\documentclass[a4paper,fleqn,usenatbib,useAMS]{mnras}

\usepackage[T1]{fontenc}
\usepackage{aecompl}

\usepackage{graphicx}	
\usepackage{amsmath}	
\usepackage{amssymb}	
\usepackage{threeparttable}
\usepackage{bm}		
\usepackage{pdflscape}	

\title[Can \ion{H}{I} 21 cm line trace the Missing Baryons in the Filamentary Structures?]{Can \ion{H}{I} 21 cm line trace the Missing Baryons in the Filamentary Structures?}
\author[T. Horii et al.]{
Toshihiro Horii,$^{}$
Shinsuke Asaba,$^{}$
Kenji Hasegawa$^{}$\thanks{E-mail: kenji.hasegawa@a.mbox.nagoya-u.ac.jp}
and Hiroyuki Tashiro$^{}$\thanks{E-mail: hiroyuki.tashiro@nagoya-u.jp}
\\
$^{}$Department of Physics, Graduate School of Science, Nagoya University, Aichi 464-8602, Japan
}

\begin{document}
\pagerange{\pageref{firstpage}--\pageref{lastpage}}
\maketitle

\begin{abstract}
A large fraction of baryons predicted from the standard cosmology has been missing observationally. 
Although previous numerical simulations have indicated that most of the
missing baryons reside in large-scale filaments in the form of Warm Hot
Intergalactic Medium~(WHIM), it is generally very difficult to detect
signatures from such a diffuse gas. 
In this work, we focus on the hyperfine transition of neutral hydrogen~(\ion{H}{I}) called 21-cm line as a tool to trace the WHIM. 
For the purpose, we first construct the map of the 21-cm signals by
using the data provided by the state-of-the-art cosmological
hydrodynamics simulation project, Illustris, in which detailed processes affecting the dynamical and thermal evolution of the WHIM are implemented. 
From the comparison with the constructed 21-cm signal map with the
expected noise level of the Square Kilometre Array phase~1 mid-frequency instrument~(SKA1-mid), we
find that the 21-cm signals from the filamentary structures at redshifts
$z=0.5-3$ are detectable with the SKA1-mid if we assume the angular resolution of $\Delta\theta\geq10$ arcmin and the observing time of $t_{\rm obs}\geq100$ hours. 
However, it also turns out that the signals mainly come from galaxies
residing in the filamentary structures and the contribution from the WHIM is too small to detect with the SKA1-mid. 
Our results suggest that about 10 times higher sensitivity than the SKA1-mid is possibly enough to detect the WHIM at $z=0.5-3$. 
\end{abstract}


\section{INTRODUCTION}\label{INTRODUCTION}

Recent various cosmological observations strongly support that
the $\Lambda$CDM~(Lambda Cold Dark Matter) model is preferred for
describing the evolution of the
Universe~\citep{2016A&A...594A..13P}. 
The present acceleration expansion revealed by observations of type Ia supernovae suggests the existence of dark energy~\citep{1997AAS...191.8504P,1998AJ....116.1009R}. 
Furthermore, numerical simulations based on the $\Lambda$CDM model succeed in explaining observed galaxy clustering in the large scale structure~\citep{2016arXiv160703155A}.
Hence, although it is a simple theoretical model,
the $\Lambda$CDM model is widely accepted as the standard model of
the Universe.
However, there are some critical cosmological problems in the
$\Lambda$CDM model to date.
While the Cosmic Microwave Background (CMB) tells us that 95\% of the total
energy density in the Universe is made of mysterious ``dark'' components~\citep{2016A&A...594A..13P},
the true natures of these dark components are totally unknown.
Even for ordinary baryonic components, although
baryonic component is responsible for $\sim$5\%~of the total energy density
in the Universe from CMB observations,
there is a problem where and in what form baryons exist~\citep{1998ApJ...503..518F}.
The baryon to photon ratio obtained from CMB observations is consistent with that 
predicted by the big bang nucleosynthesis with local measurements of the light elements~\citep{2013arXiv1307.6955C}.
However, only 10\% of the predicted amount of baryons are found as stars and gas clouds 
in galaxies and galaxy clusters~\citep{1999MNRAS.309..923S,2004ApJ...616..643F}.
Therefore, other baryons are expected to be outside galaxies and this is known as the ``missing baryon''. 
The dynamical evolution of the dark matter density fluctuations naturally leads to the formation of large scale structures, so-called cosmic webs. 
Recent cosmological structure formation simulations have shown that the missing
baryons are in the form of shock heated warm/hot intergalactic gas
called WHIM whose temperature correspond to $10^5$--$10^7$~K \citep[e.g.][]{Cen1999, Dave2001, Yoshikawa2003, Bregman2007}.
Therefore, surveys of the WHIM in large-scale filamentary structures provide the key to access the missing baryons.

Observations of the metal absorption lines imprinted on high-redshift QSO
spectra play important roles to survey the WHIM.
\citet{Tripp2001} have firstly confirmed that the predicted hot diffuse
gas actually exists at low redshifts, 
using high-resolution spectra of QSOs.
Subsequent observations including Ly-$\alpha$ surveys
revealed that roughly 30\% of baryons exists as the WHIM
whose temperature is in the order of $10^5$~K~\citep{Nicastro2008,Shull2012}.
Although these UV and optical observations are too low resolution to
point out the location and topology of the WHIM,
recent X-ray observations can trace filamentary structures of the WHIM. 
\citet{Kull1999} reported the existence of
inter-cluster filament structure of gas in the central region of Shapley supercluster
by X-ray observations. 
In \cite{Werner2008}, the authors found X-ray emission from 
the filamentary structure between the galaxy clusters, Abell222 and Abell223.
The observation suggests that, in this filament region, the gas temperature and
density are as high as
the densest and hottest WHIM in simulations.
More recently, \cite{Nature2015} have finally found hot gas associated
with the filaments surrounding the galaxy cluster Abell 2744 via the
X-ray observation whose spatial resolution is high enough to trace the filamentary structures. 
Although such observations are incrementally revealing the existence of
the missing baryons,
still 30--40\% of baryons are missing~\citep{2004ApJ...616..643F,Shull2012}.
In particular, it is thought that the WHIM with temperature $> 10^6$~K has not been fully explored yet.

Attempts to detect the baryons which reside in the large-scale filaments have been hitherto limited to approaches with X-ray and UV/opt observations. 
However forthcoming radio telescopes are promising to detect the filaments. 
For instance the contribution of the thermal and kinematic Sunyaev-Zel'dovich effect through the WHIM is discussed in~\citet{2006ApJ...643....1A,2008ApJ...674L..61A,2013A&A...550A.134P}.
Furthermore, if the filamentary structures are magnetized, the large-scale filamentary structures can be traced by the synchrotron radiation~\citep{2012MNRAS.423.2325A,2014MNRAS.445.3706V} and the Faraday tomography \citep{Akahori14} with a next generation radio interferometer Square Kilometre Array~(SKA). 

Another promising approach to cast light on the baryons that reside in the large-scale filaments is to observe the hyper fine structure line of neutral hydrogen, i.e. \ion{H}{I} 21-cm line. 
It is true that the Universe is mostly ionized at $z<6$, but the SKA is
expected to have an enough potential to detect weak \ion{H}{I} 21-cm
emission lines from residual neutral hydrogen in the filaments. 
\citet{Takeuchi2014} have investigated the detectability of such \ion{H}{I}
21-cm signals from the filaments by performing an $N$-body simulation. 
To evaluate the 21-cm signals from filaments, they have assumed that the thermal and ionization evolution of gas in
filaments are determined by background UV radiation and the Hubble expansion.
However, as shown in cosmological hydrodynamics simulations, 
the shock heating during the structure formation and feedback effects from galaxies, which cannot be taken into account in $N$-body simulations, play important roles in determining the state of the WHIM in filamentary structures.
The shock heating can heat up the gas residing in filamentary regions up to $10^{5}-10^{7}$K~\citep{Dave2001,Bregman2007}, and outflowing gas from galaxies could also provide an impact on the evolution of gas. 
This underestimation of the gas temperature very likely results in
overestimation of the neutral fraction of gas in the filaments
due to the lack of the collisional ionization process. 
Furthermore, it can be expected that galaxies in filamentary structures also significantly contribute to 21-cm signals because they can host substantial neutral hydrogen as overdense clouds.
Therefore, the detectability of \ion{H}{I} 21-cm signals from the filamentary structures or the WHIM in them is still open to debate. 

In this paper, we revisit the detectability of large-scale filamentary structures at high redshifts
through the \ion{H}{I} 21-cm signals with the SKA. 
In particular, we discuss the potential of 21-cm observations to probe
the WHIM in filamentary structures to resolve the missing baryon problem.
For this purpose, we use the publicly opened data from the Illustris simulation in which detailed baryonic processes are taken into account \citep{Genel2014,Vogelsberger2014}\footnote{\url{http://www.illustris-project.org/}}. 
In this simulation, the cosmological parameters are adopted as
$\Omega_{\rm b}=0.0456$, $\Omega_0=0.273$, $\Omega_{\Lambda}=0.727$,
$H_{0}=70.4\rm{km/s/Mpc}$, $\sigma_8=0.809$ and $n_{\rm s}=0.963$, and we also
use these parameters to calculate the 21-cm signals.

This paper is organized as follows. 
In \S~\ref{METHOD}, we show how we evaluate the expected \ion{H}{I}
21-cm signals from the simulation data. 
A brief description of the Illustris simulation is also presented in this section. 
We show our results in \S~\ref{Detectability of HI 21-cm line}. 
\S~\ref{DISCUSSION} is devoted for discussions. 
Finally we give a summary of this paper in \S~\ref{CONCLUSIONS}

\section{construction of the 21-cm signal map}\label{METHOD}

The measurement of~\ion{H}{I} 21-cm signals is a powerful tool for
understanding the state of baryonic gas.
In this section, we briefly review the physics of the \ion{H}{I} 21-cm
signals and describe how we evaluate the \ion{H}{I} 21-cm signals from the
Illustris simulation data. 

\subsection{Differential brightness temperature}\label{Differential brightness temperature} 

When we observe 21-cm signals, the absolute value of these brightness
temperature cannot be measured
directly because of the existence of the CMB over the full frequency range.
Therefore, the measured quantity is the difference from the CMB
brightness temperature called the differential brightness temperature.
The observable differential brightness temperature of 21-cm signals from
the IGM is given by~\citep{1997ApJ...475..429M,Furlanetto2006}
\begin{align}
	\delta T_{\rm{b}}\equiv T_{\rm{b}}-T_{\rm{CMB}}=\frac{[T_{\rm s}-T_{\rm{CMB}}(z)](1-e^{-	\tau})}{1+z},
	\label{deltatb}
\end{align}
where $T_{\rm CMB}$, $T_{\rm{b}}$ and ${T_{\rm s}}$
are the CMB temperature, the brightness temperature of \ion{H}{I} 21-cm
line and its spin temperature, respectively. 
The 21-cm optical depth~$\tau$ of the IGM in Eq.~(\ref{deltatb}) is given by 
\begin{align}
	\tau=\frac{g_{1}}{g_{0}+g_{1}}\frac{h_{\rm{p}} c^{3} A_{10}}{8\pi \nu_{10}^{2} 
	k_{\rm{B}}}\frac{n_{\ion{H}{I}}}{T_{\rm s}}\frac{1}{(1+z)(dv_{||}/dr_{||})}, 
	\label{tau1}
\end{align}
where the subscripts $0$ and $1$ denote the hyperfine ground and excited states, respectively. 
Here, $g$ is the statistical weight of neutral hydrogen, $A_{10}$ is the
Einstein A coefficient for the transition, $n_{\ion{H}{I}}$ is the
number density of neutral hydrogen,
$\nu_{10}$ is the frequency corresponding to the energy difference between these states, $h_{\rm{p}}$ is the Planck constant, $c$ is the speed of light, and $k_{\rm{B}}$ is the Boltzmann constant.
In Eq.~(\ref{deltatb}), $dv_{||}/dr_{||}$ is the velocity gradient along
the line-of-sight that usually involves the effects of the Hubble
expansion, the peculiar velocity and the thermal velocity of the gas. 
In this paper,
we simply take into account the effect of the Hubble expansion,
$(1+z)(dv_{||}/dr_{||}) \approx H(z)$\footnote{In particular, the thermal velocity and the peculiar velocity
become important at overdense regions. We find that the velocity effects
could suppress the signal amplitude almost by
half. However this suppression of the signals does not modified our conclusion on the detectability of the filaments.}. 
As a result, we can rewrite Eq.~(\ref{tau1}) into 
\begin{align}
	\tau \approx \frac{g_{1}}{g_{0}+g_{1}}\frac{h_{\rm{p}} c^{3} A_{10}}{8\pi \nu_{10}^{2} 	
	k_{\rm{B}}}\frac{n_{\ion{H}{I}}}{T_{\rm s}}\frac{1}{H(z)}, 
	\label{tau2}
\end{align}
When we consider \ion {H}{I} 21-cm signal from the diffuse IGM, the
optically thin approximation, $\tau \ll 1$, generally holds. 
With Eq.~(\ref{tau2}), Eq.~(\ref{deltatb}) is therefore approximated as 
\begin{align}
	\delta T_{\rm b}&\simeq \frac{T_{\rm{s}}-T_{\rm{CMB}}(z)}{1+z}\tau \nonumber \\
	& \simeq\frac{3}{32\pi}\frac{h_{\rm{p}}c^{3}A_{10}}{k_{\rm{B}}\nu_{10}^{2}}\frac{1}{1+z}	
	x_{\ion{H}{I}} \frac{n_{\rm{H}}}{H(z)} \times \biggl( 1-\frac{T_{\rm{CMB}}(z)}{T_{\rm{s}}} \biggl).
	\label{dtb2}
\end{align}
where $x_{\ion{H}{I}}$ is the neutral hydrogen fraction and $n_{\rm{H}}$ is the total hydrogen number density. 
Eq.~(\ref{dtb2}) tells us that we can estimate the differential
brightness temperature
once we know
$T_{\rm s}$ and $x_{\ion{H}{I}}$. 

\subsection{Spin temperature}\label{Spin temperature}
There are three physical processes contributing the spin temperature. 
Namely, the excitation and de-excitation by the CMB photons, the
collisions with electrons and other atoms, and the interaction with
background Lyman-$\alpha$ photons (known as the Wouthuysen-Field effect, hereafter the WF effect). 
Hence, the spin temperature is given by~\citep{1959ApJ...129..536F}
\begin{align}
	T_{\rm s}^{-1}=\frac{T_{\rm{CMB}}^{-1}+x_{\rm c}T_{\rm{K}}^{-1}+x_{\alpha}
	T_{\alpha}	^{-1}}{1+x_{\rm c}+x_{\alpha}} , \label{spin}
\end{align}
where $x_{\rm c}$ and $x_{\alpha}$ are the coupling coefficients for the collisional and WF processes, respectively. 
Besides, $T_{\rm{K}}$ is the kinetic temperature of the gas,
$T_{\alpha}$ is the color temperature in the vicinity of the Ly-$\alpha$ frequency.

We can write the collisional coupling coefficient $x_{\rm c}$ as
\begin{align}
	x_{\rm c}=\frac{T_{\ast}}{A_{10}T_{\rm{CMB}}}(C_{\ion{H}{I}}+C_{\rm{p}}+C_{\rm{e}}),
\label{eq:xc-def}
\end{align}
where $T_{\ast}$ is the temperature corresponding to the hyperfine transition and is defined as 
$T_{\ast}\equiv h_{\rm{p}} \nu_{10}/k_{\rm{B}}$. 
In Eq.~(\ref{eq:xc-def}),
$C_{\ion{H}{I}}$, $C_{\rm{p}}$, and $C_{\rm{e}}$ are respectively the collisional de-excitation rates by neutral hydrogen, protons, and electrons. 
In this work, we use the fitting formula shown by \citet{Kuhlen2006}
for these de-excitation rates, $C_{\ion{H}{I}}$, $C_{\rm{p}}$, and
$C_{\rm{e}}$.
The collisional de-excitation rate by neutral hydrogen $C_{\ion{H}{I}}$ is represented as 
\begin{align}
	C_{\ion{H}{I}}(T_{\rm{K}})=n_{\ion{H}{I}}\gamma_{\ion{H}{I}}(T_{\rm{K}}). \label{CH}
\end{align}
In the $T_{\rm{K}}<10^{3}~\rm{K}$ regime, $\gamma_{\ion{H}{I}}(T_{\rm K})$ is described as 
\begin{align}
	\gamma_{\ion{H}{I}}(T_{\rm{K}})=\gamma_{\ion{H}{I}, L}(T_{\rm{K}})=3.1\times10^{-11}\times T_{\rm{K}}^{0.357}\exp(-32/T_{\rm{K}})~
	\rm{cm}^{3}\rm{s}^{-1} , \label{gamma}
\end{align}
otherwise, $\gamma_{\ion{H}{I}}(T_{\rm{K}})$ is written as $4\gamma_{\ion{H}{I}, L}(T_{\rm{K}} /3)$. 
Using the electron number density $n_{\rm e}$, we show the collisional de-excitation rate by electron $C_{\rm{e}}$ as 
\begin{align}
	C_{\rm{e}}(T_{\rm{K}})=n_{\rm{e}}\gamma_{\rm{e}}(T_{\rm{K}}),  \label{Ce}
\end{align}
letting $\gamma_{\rm e}(T_{\rm K})$ be 
\begin{align}
	\log \biggl(\frac{\gamma_{\rm{e}}}{\rm{cm^{3}s^{-1}}}\biggl)=-9.607+0.5\log(T_{\rm{K}})\exp\biggl[-\frac{(\log T_{\rm{K}})^{4.5}}{1800}\biggl]. \nonumber
\end{align}
Here, we set $\gamma_{\rm{e}}(T_{\rm{K}}>10^{4}~\rm{K})=\gamma_{\rm{e}}(T_{\rm{K}}=10^{4}~\rm{K})$. 
Finally, the de-excitation rate by protons is given by 
\begin{align}
	C_{\rm{p}}(T_{\rm{K}})=n_{\rm{p}}\gamma_{\rm{p}}(T_{\rm{K}}), \label{Cp}
\end{align}
where $n_{\rm{p}}$ is the proton number density and $\gamma_{\rm{p}}$ can be written as $\gamma_{\rm{p}}(T_{\rm{K}})=3.2\gamma_{\ion{H}{I}} (T_{\rm{K}})$. 

The coupling coefficient for the WF effect $x_{\alpha}$ obeys 
\begin{align}
	x_{\alpha}=1.81\times 10^{11} (1+z)^{-1} S_{\alpha} J_{\alpha}/h\nu_\alpha, 
\end{align}
where $S_{\alpha}$ is the scattering amplitude factor, $\nu_\alpha$ is
the Lyman-$\alpha$ frequency,  
$J_{\alpha}$ is the Lyman-$\alpha$ mean intensity. 
According to \citet{Furlanetto2006}, 
$S_{\alpha}$ is described as 
\begin{align}
	S_{\alpha} \sim \exp \biggl[ -0.803 T_{\rm{K}}^{-2/3} \biggl( \frac{10^{-6}}{\gamma} \biggl)^{1/3} \biggl], 
\end{align}
where $\gamma$ is the Sobolev parameter.
It has been known that the Sobolev parameter $\gamma$ corresponds to the inverse of the Gunn-Peterson optical depth.
Thus, we can write  
\begin{align}
	\gamma =\tau_{\rm{GP}}^{-1}=\frac{H(z)\nu_{\alpha}}{\chi_{\alpha}n_{\ion{H}{I}}(z)c}, 
\end{align}
where , $\nu_{\alpha}$ is the $\rm Ly_{\alpha}$ frequency, i.e. $2.47\times 10^{15}~\rm{Hz}$. 
Besides, $\chi_{\alpha}$ is defined as 
\begin{equation}
	\chi_{\alpha}\equiv (\pi e^{2}/m_{\rm{e}}c)f_{\alpha},
\end{equation}
where $m_{\rm e}$ is the electron mass, $f_{\alpha}$ is the oscillator strength for $\rm Ly_{\alpha}$ transition, and hence $f_{\alpha}$. 
For the Lyman-$\alpha$ background mean intensity $J_{\alpha}$, 
we use the values in Table~\ref{tab:mean intensity} from~\citet{HM2012}.

\begin{table}
\caption{Background $\rm Ly_{\alpha}$ mean intensity $J_{\alpha}$ from \citet{HM2012}}
\label{tab:mean intensity}
\begin{tabular}{lcc}
\hline
Redshift & Mean intensity [$\rm{erg}\,\rm{cm^{-2}\, s^{-1}\, Hz^{-1}\, sr^{-1} }$]\\
\hline
$0.5$ & $9.59\times10^{-21}$\\
$1.0$ & $2.89\times10^{-20}$\\
$2.0$ & $7.25\times10^{-20}$\\
$3.0$ & $6.31\times10^{-20}$\\
\hline
\end{tabular}
\label{tab:mean intensity}
\end{table}

\subsection{Data: Illustris simulation}\label{Illustris simulation}

As described above, the differential brightness temperature of 21-cm line can be evaluated from the gas temperature and the number densities of electrons, protons, and neutral hydrogen. 
In order to obtain the map of the differential brightness temperature,
we use the results of the state-of-the-art cosmological hydrodynamics
simulation project called Illustris in which the thermal and dynamical evolution of baryons is solved with a moving-mesh code AREPO~\citep{Springel2010, Genel2014,Vogelsberger2014}. 
As already mentioned in \S~\ref{INTRODUCTION}, it is essential to consider baryonic processes, such as radiative cooling and shock heating, for evaluating \ion{H}{I} 21-cm signal from WHIM. 
The Illustris simulation appropriately involves these baryonic processes as well as feedback processes driven by supernovae and active galactic nuclei to make simulated galaxies reproduce the observed luminosity function at the present-day \citet{Vogelsberger2013,Torrey2014}. 

Since the thermal and chemical evolution is solved in the Illustris
simulation, the publicly available data set contains the information
required for evaluating the differential brightness temperature,
i.e. neutral hydrogen number density, electron number density, proton
number density, and gas temperature, except the $\rm Ly_\alpha$
background mean intensity $J_{\alpha}$. 
As described in \S~\ref{Spin temperature}, we refer to \citet{HM2012} for $J_{\alpha}$

We use the data of Illustris simulation-3; 
The simulation volume is $106.5$Mpc on a side, and $2\times455^3$ gaseous cells and DM particles are distributed in the volume.
The resultant mass resolutions respectively correspond to $8.1\times 10^7~{\rm M_{\odot}}$ for gas and $4.8\times 10^8~{\rm M_{\odot}}$ for DM \citep{Nelson2015}.  

\subsection{Map of 21-cm signals from the Illustris data}

\begin{figure}
  \begin{tabular}{cc}
  \begin{minipage}[b]{0.5\linewidth}
    \centering
    \includegraphics[width=\columnwidth]{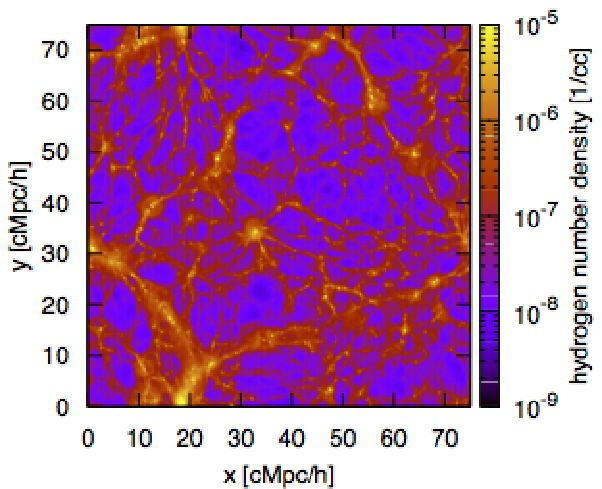}
  \end{minipage}
  \begin{minipage}[b]{0.5\linewidth}
    \centering
    \includegraphics[width=\columnwidth]{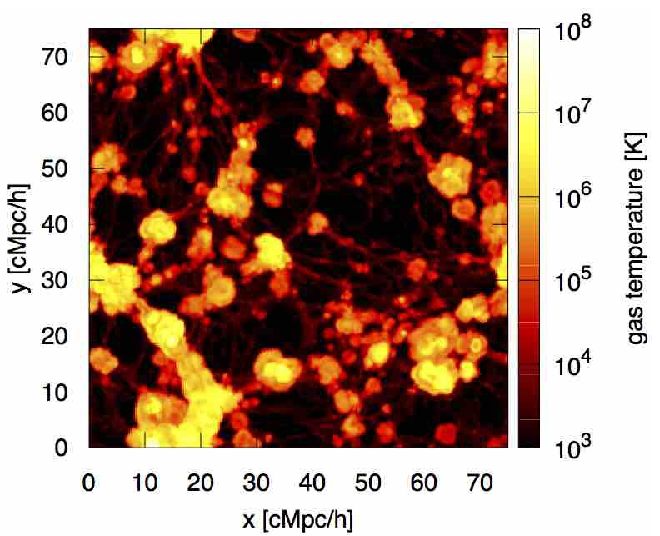}
  \end{minipage}
    \end{tabular}
   \begin{tabular}{cc}
  \begin{minipage}[b]{0.5\linewidth}
    \centering
    \includegraphics[width=\columnwidth]{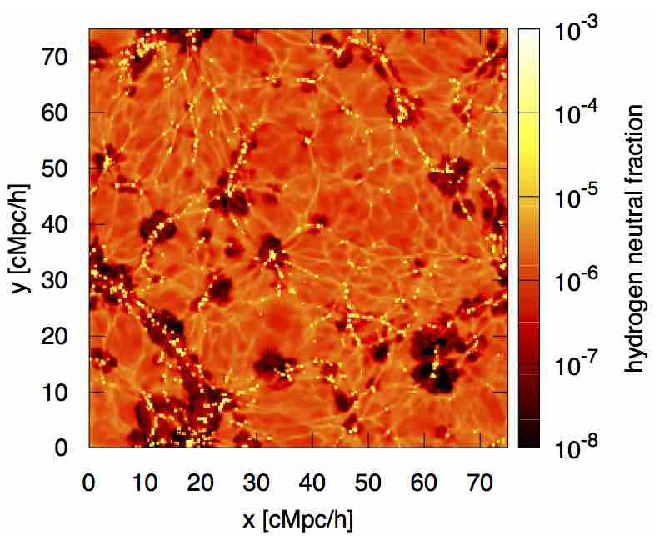}
  \end{minipage}
  \begin{minipage}[b]{0.5\linewidth}
    \centering
    \includegraphics[width=\columnwidth]{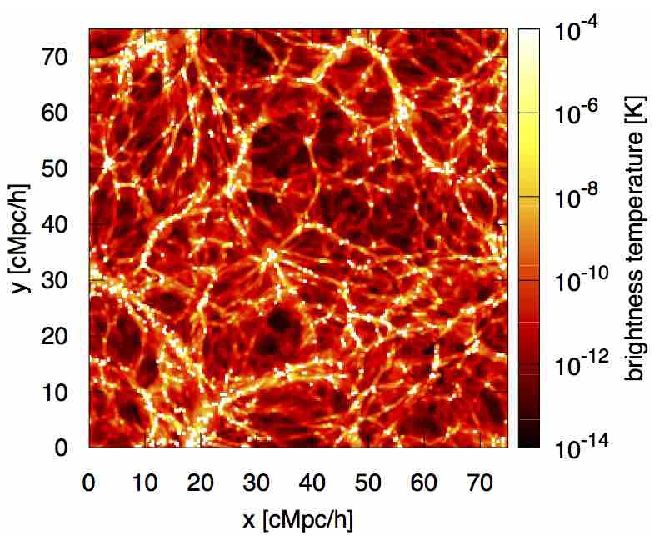}
  \end{minipage}
  \end{tabular}
  \caption{Maps of the hydrogen number density~(top left), the gas temperature~(top right), 
 the neutral fraction~(bottom left), and the differential brightness temperature~(bottom right) at
 the present epoch. The size is 75~Mpc/h on a side with a thickness of 1.5~Mpc/h. }\label{fig:z0}
\end{figure}

We construct the maps of the brightness temperature by following the
method described in \S~\ref{METHOD} and show the resultant maps in Fig.~\ref{fig:z0}.
The left top, right top, left bottom, and right bottom panels in Fig.~\ref{fig:z0}
respectively show the maps of the total hydrogen number density~($n_{\rm H}$), the gas temperature~($T_{\rm K}$), the neutral fraction~($x_{\ion{H}{I}}$), and the differential brightness temperature~($\delta T_{\rm b}$). 

In the left top panel describing the distribution of hydrogen number
density, we can identify many filamentary structures. 
As previous numerical simulations have shown, the gas in the filaments
are well heated by shocks up to $\sim10^5$-$10^6$~K in the right top panel. 
In the filaments, there exist more overdense regions in which the temperature is higher than around them and reaches $>10^6$~K. 
These regions are close to galaxies, and thus the efficient heating due to strong outflows from galaxies is remarkable. 

As shown in the left bottom panel, the neutral hydrogen fraction
in the filament structures are very small due to the high collisional ionization rate. 
The brightness temperature depends on the neutral hydrogen
density, namely, the combination of the neutral hydrogen fraction and
the gas density.
Therefore,
in spite of the low neutral hydrogen fraction, the higher density of gas makes the resultant brightness temperatures higher in the filamentary structures than those in the lower density regions.
Therefore, we can conclude that the distribution of the 21-cm brightness temperature well traces the filament structures as seen in the left bottom in Fig.~\ref{fig:z0}. 

\begin{figure}
    \begin{tabular}{cc}
 \begin{minipage}[b]{0.5\linewidth}
    \centering
    \includegraphics[width=\columnwidth]{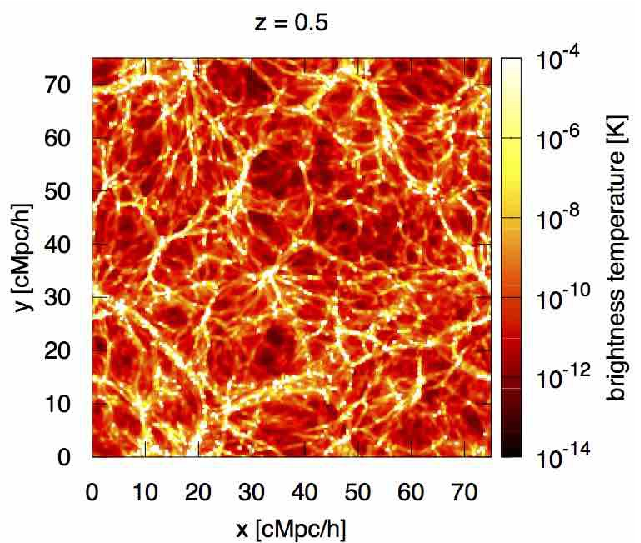}
 \end{minipage}
  \begin{minipage}[b]{0.5\linewidth}
    \centering
    \includegraphics[width=\columnwidth]{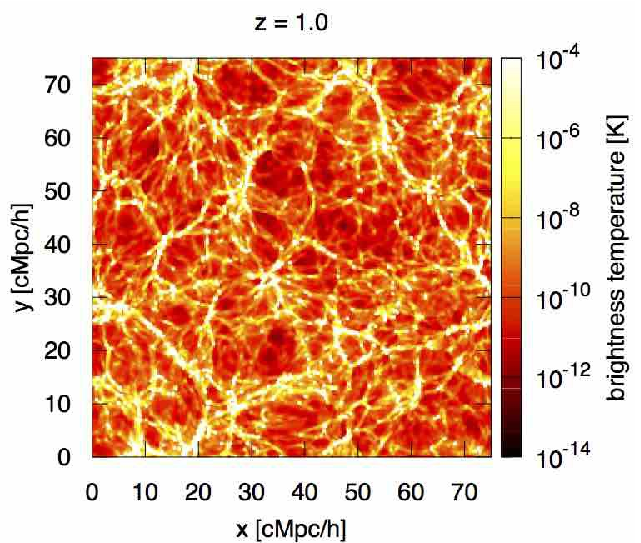}
  \end{minipage}
     \end{tabular}
         \begin{tabular}{cc}
  \begin{minipage}[b]{0.5\linewidth}
    \centering
    \includegraphics[width=\columnwidth]{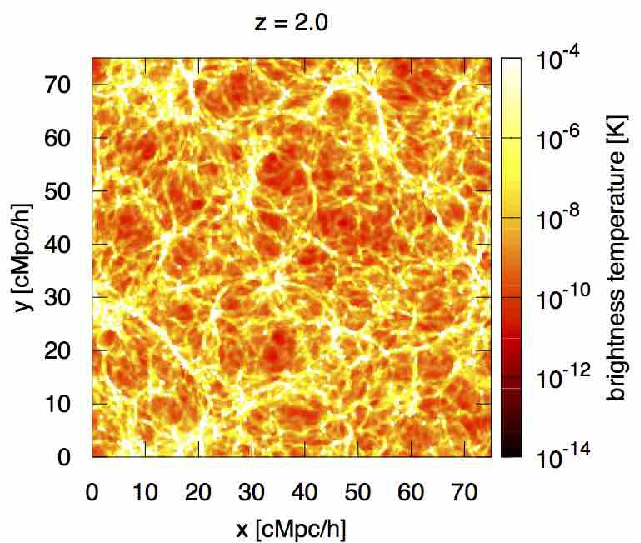}
  \end{minipage}
  \begin{minipage}[b]{0.5\linewidth}
    \centering
    \includegraphics[width=\columnwidth]{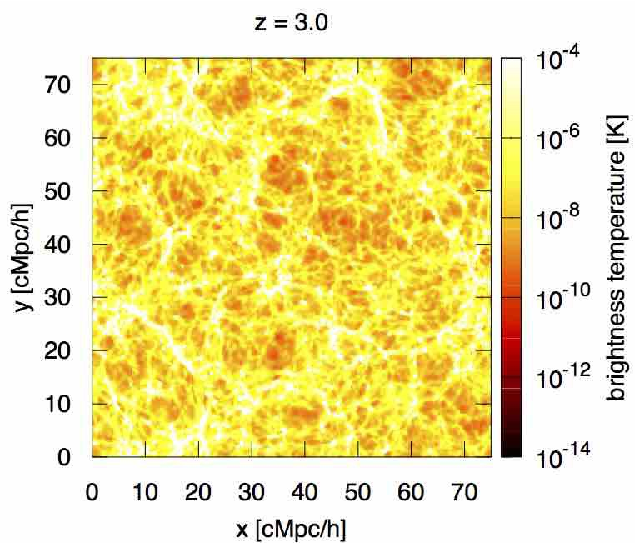}
  \end{minipage}
	  \end{tabular}
  \caption{Similar to the right bottom panel in Fig.~\ref{fig:z0}, but for $z=0.5$~(left top), $z=1.0$~(right top), $z=2.0$~(left bottom), $z=3.0$~(right bottom). }\label{fig:deltaTb}
\end{figure}

We also show the redshift evolution of the brightness temperature maps for $z=0.5$, 1.0, 2.0 and 3.0 in Fig.~\ref{fig:deltaTb}. 
From Fig.~\ref{fig:deltaTb}, we can see that the distributions of the brightness temperature nicely trace the filaments especially at lower redshifts.
The gas density and the neutral fraction also rise as the redshift becomes higher. 
This fact leads to the increase in the signal amplitude, although identifying the filaments on the map becomes difficult due to the weak contrast of the brightness temperature. 

\section{Detectability of \ion{H}{I} 21-cm line}\label{Detectability of HI 21-cm line} 

\begin{figure}
 \includegraphics[width=\columnwidth]{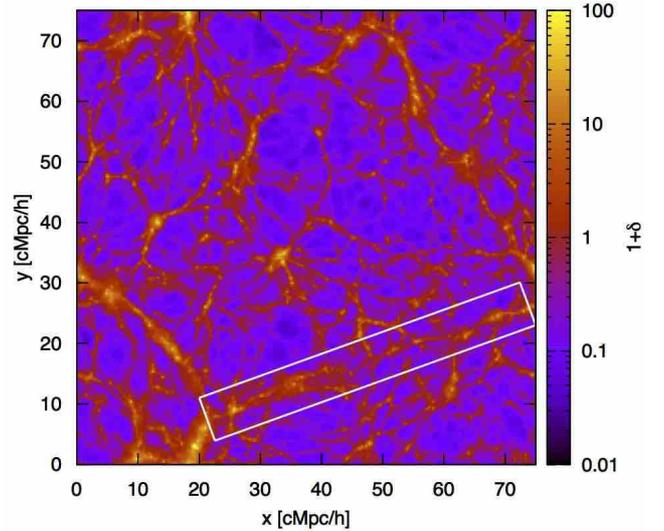}
 \caption{Map of the hydrogen number density. The size is 75 Mpc/h on a side with a thickness of 1.5 Mpc/h. The region surrounded by a white square is defined as the filament structure.}
 \label{fig:overdensity256}
\end{figure}

As shown above, the measurement of 21-cm signals has a potential to probe filaments. 
In this section, we evaluate the detectability of the signals from
filamentary structures with the SKA. 
For the discussion, we focus on a filamentary structure surrounded by the white square in Fig.~\ref{fig:overdensity256}.

\begin{figure}
 \includegraphics[width=\columnwidth]{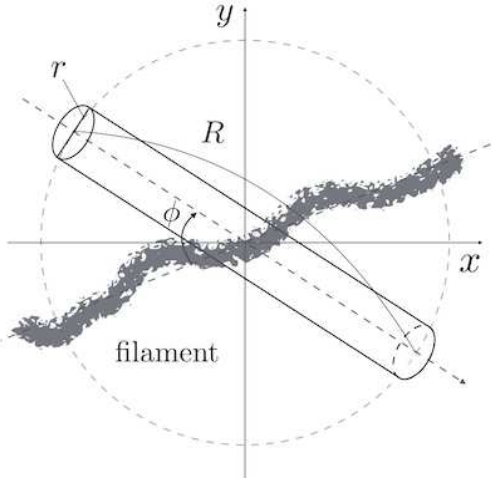}
 \caption{Relation between the observation cylinder along  the line of sight and the filament structure that we select. $R$ and $r$ respectively represent the length and the width of the cylinder corresponding to the bandwidth and the angular resolution of an interferometer telescope. Also, $\phi$ represents the angle between axes of the cylinder and the filament.}
 \label{fig:filament}
\end{figure}

\begin{figure*}
  \begin{minipage}[b]{0.45\textwidth}
\centering
$\Delta \theta = 3$ arcmin
\includegraphics[width=\linewidth]{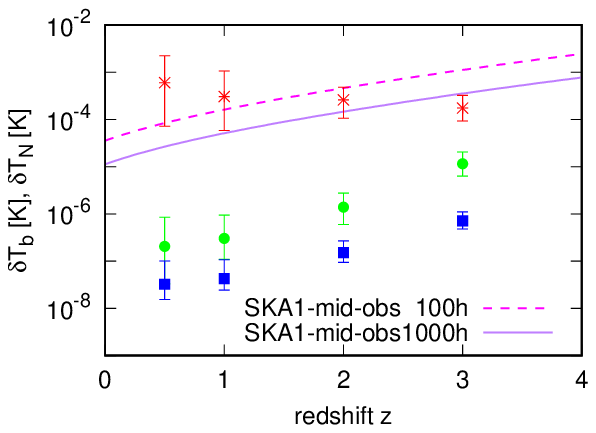}
  \end{minipage}
  \begin{minipage}[b]{0.45\textwidth}
 \centering
$\Delta \theta = 10$ arcmin
 \includegraphics[width=\linewidth]{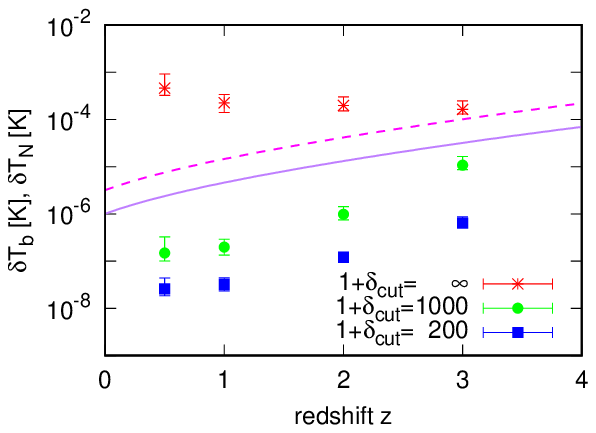}
  \end{minipage}\\
  \begin{minipage}[b]{0.45\textwidth}
 \centering
$\Delta \theta = 20$ arcmin
 \includegraphics[width=\linewidth]{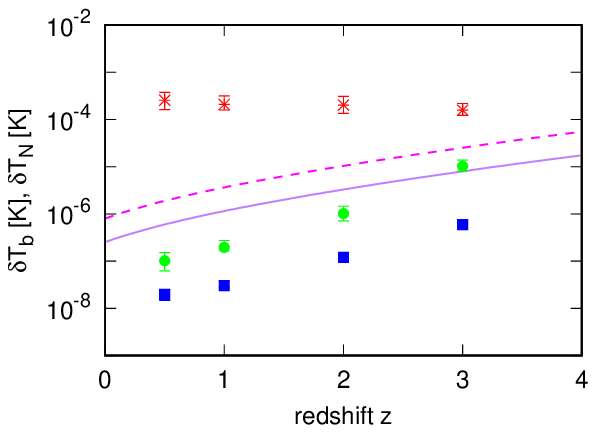}
  \end{minipage}
  \begin{minipage}[b]{0.45\textwidth}
 \centering
$\Delta \theta = 30$ arcmin
 \includegraphics[width=\linewidth]{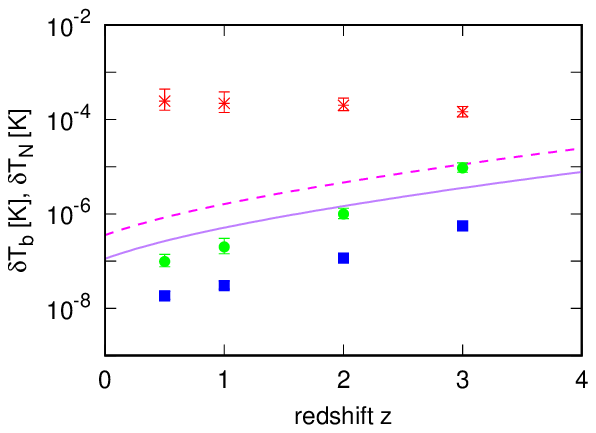}
  \end{minipage}
  \caption{Redshift evolution of the 21-cm signal and the noise
 of the future interferometer telescope, SKA1-mid. The solid line and
 the dashed line are the noise for the total observing time
 $t_{\rm{obs}} = 1,000~\rm{hours}$ and $t_{\rm{obs}} = 100~\rm{hours}$.
The red asterisks represents the signals including galaxies, the CGM and the
 WHIM in the filament. 
 The green filled circles and the blue filled squares indicate the signals
 calculated without the overdense regions with the density
 contrast larger than $\delta_c>1,000$~(the CGM and the WHIM) and
 $>200$~(only the WHIM), respectively. 
 The bars represent the range of signal between the maximum and minimum values.
 Here, we set the bandwidth $\Delta \nu = 3~\rm{MHz}$ and use $\Delta \theta = 3$ arcmin (left top), $\Delta \theta = 10$ arcmin (right top), $\Delta \theta = 20$ arcmin (left bottom) and $\Delta \theta = 30$ arcmin (right bottom), respectively, as the angular resolution. }
 \label{fig:angreso}
 \end{figure*}

The observed signal depends on the angular resolution $\Delta \theta$ and the
bandwidth $\Delta \nu$ of an interferometer telescope.
To take $\Delta \theta$ and $\Delta \nu$ into account, we follow~\citet{Takeuchi2014} and consider an observation cylinder aligned along the line of sight shown in Fig.~\ref{fig:filament}.
We evaluate the observed signal from the filamentary structure averaged over the observation cylinder. 
In this configuration, the width $r$ and length $R$ of the cylinder correspond to
$\Delta \theta$ and $\Delta \nu$, respectively. 
Therefore, the signal also depends on how the observation cylinder intersects the
filament structure. We introduce the parameter $\phi$ which describes
the angle between the axes of the filament structure and the observation cylinder in Fig.~\ref{fig:filament}.
When $\phi =0$, the axis of the cylinder is completely parallel to
the long axis of the filament structure.
Laying a cylinder on a 21-cm map, we calculate the amplitude of the signal for each $\phi$ between -90 and 90~degree.
The redshift evolution of the 21-cm signal is shown as red asterisks in Fig.~\ref{fig:angreso}.
Here we set $\Delta \nu =3$~MHz and show the expected signals for $\Delta \theta =$3, 10, 20 and 30 arcmin in the left upper, right upper, left lower and right lower panels, respectively.
Since the signal amplitude depends on $\Delta \theta$, the red asterisks in the figure show the averaged values and the red bars represent the maximum and minimum values between $\phi=$-90 and 90~degree. 
The signals reach the maximum when $\phi = 0$.

Fig.~\ref{fig:angreso} also shows that the amplitude of the signals increases as the observation redshift increases. 
As mentioned above, the density contrast is small in the filamentary structures at high redshifts, because the filamentary structures do not evolve well yet. 
However the baryon density itself is high at high redshifts and we find that the neutral fraction also increases. Therefore, the 21-cm signal increases with increasing the observation redshift. 

To evaluate the detectability, we compare these signals with the noise level of the Square Kilometre Array phase~1 mid-frequency instrument~(SKA1-mid).
According to~\citet{Furlanetto2006}, the noise level of an
interferometer is given in terms of the brightness temperature by
\begin{align}
\delta T_{\rm{N}}(\lambda)=
 \frac{\lambda^{2}/\Delta
 \theta^{2}}{A_{\rm{eff}}}\frac{T_{\rm{sys}}}{\sqrt{\Delta \nu \,
 t_{\rm{obs}}}}, \label{eq:noise_eq}
\end{align}
where $\Delta \theta$ is the beam width, $\Delta \nu$ is the bandwidth,
$A_{\rm{eff}}$ is the effective collecting are, $T_{\rm sys}$ is the
system temperature and $t_{\rm obs}$ is the total observing time.

The system temperature can be divided into two components as
\begin{align}
T_{\rm{sys}} = T_{\rm{rec}} +T_{\rm{sky}},
\end{align}
where $T_{\rm rec}$ is for
the instrumental noise temperature and $T_{\rm{sky}}$ represents the sky temperature.
Here, as described below, we set $T_{\rm rec} =30$~K.
On the other hand, for $T_{\rm sky}$
we adopt the sky temperature at high Galactic latitude, in  which the
foreground emission is minimum in the sky.
The sky temperature at this region is roughly approximated as
\begin{align}
T_{\rm{sky}}\sim 180\biggl(\frac{180~\rm{MHz}}{\nu}\biggl)^{2.6}~\rm{K} .
\end{align}
There exists the difference of the frequency dependence
between $T_{\rm rec}$ and $T_{\rm sky}$. 
The noise due to the sky temperature 
dominates at low frequencies~($\nu \lesssim 150~\rm{MHz}$),
while the instrumental noise makes a main contribution at high
frequencies.
Therefore, the resultant noise brightness temperature 
can be approximated as 
\begin{align}
\delta T_{\rm{N}}(\nu)\simeq
 {7.36}
 \biggl( \frac{10^{5}~\rm{m}^{2}}{A_{\rm{eff}}} \biggl) \biggl(
 \frac{1'}{\Delta \theta} \biggl)^{2}
 \biggl( \frac{\rm{MHz}}{\Delta \nu} \frac{100~\rm{h}}{t_{\rm{obs}}}\biggl)^{1/2}
 \nonumber \\
 \quad \times \biggl( \frac{1+z}{1} \biggl)^{4.6} \biggl( \frac{1420~\rm{MHz}}{\nu_{10}} \biggl)^{4.6}
 \mu \rm{K}, 
\end{align}
for $\nu < 150\rm{MHz}$.
On the other hand, for higher frequencies, $\nu > 150~\rm{MHz}$, the noise
signal is 
\begin{align}
\delta T_{\rm{N}}(\nu)\simeq
{264}
 \biggl( \frac{T_{\rm{rec}}}{30~\rm{K}} \biggl) \biggl(
 \frac{10^{5}~\rm{m}^{2}}{A_{\rm{eff}}} \biggl) \biggl( \frac{1'}{\Delta
 \theta} \biggl)^{2} 
 \biggl( \frac{\rm{MHz}}{\Delta \nu} \frac{100~\rm{h}}{t_{\rm{obs}}} \biggl)^{1/2}
 \nonumber  \\
\quad \times \biggl( \frac{1+z}{1} \biggl)^{2.0}
 \biggl( \frac{1420~\rm{MHz}}{\nu_{10}} \biggl)^{2.0}
\mu \rm{K}.
\end{align}

Currently, the SKA1-mid is designed to be $A_{\rm{eff}}=48,900~\rm{m^{2}}$, $\Delta\nu=3~\rm{MHz}$ and
$T_{\rm{rec}}=30~\rm{K}$.
We calculate the noise signals from Eq.~(\ref{eq:noise_eq}) for
different angular resolutions and observing time
and plot them in Fig.~\ref{fig:angreso}.
The solid lines represent the noise for the observing time of $t_{\rm obs} = 1,000$~hours and the dashed lines for $t_{\rm obs} = 100$~hours.
As shown in Eq.~(\ref{eq:noise_eq}), the noise  increases as the
angular resolution becomes small. In the case with the small
angular resolution of $\Delta \theta =3$~arcmin, the noise is comparable
with the 21-cm signals from the filament structure at lower redshifts and dominates the 21-cm
signals at high redshifts.
Therefore, we conclude that it is difficult to probe the filamentary structures by utilizing the SKA1-mid with $\Delta \theta =3$~arcmin.
However, the observation with $\Delta \theta =10$~arcmin and $t_{\rm obs} = 1,000$~hours
provides more than 2 $\sigma$ detection of the 21-cm signal even at $z=3$. 
At lower redshifts $z<3$, the 21-cm signals completely dominate the
noise levels for both $t_{\rm obs} =100$ and 1,000~hours.
Although the observation with $\Delta
\theta=20$~arcmin can easily detect the signals from the filaments up to the
redshift $z=3$, this resolution is not so enough to map the detailed filamentary structures.
 
\section{DISCUSSION}\label{DISCUSSION}

\subsection{Contribution of galaxies and IGM to the 21-cm signal}

In \S~\ref{Detectability of HI 21-cm line}, we have shown that the SKA1-mid is capable of probing the filamentary structures at $z<3$. 
However, it is still not clear what is the dominant contributor of the 21-cm signal and whether we can detect the diffuse IGM component with the SKA1-mid. 

It is expected that galaxies in a filamentary structure can provide a significant contribution to the 21-cm signal from the filament, because galaxies hold a large amount of neutral hydrogen.
To quantify the contribution from galaxies in the Illustris simulation data, we plot $x_{\ion{H}{I}}$ as a function of the number density of hydrogen nucleous in Fig.~\ref{fig:nxT}. 
The color in the figure indicates the gas temperature. 
In the simulation, galaxies generally locate in overdense regions, e.g. $n_{\rm H}>10^{-4}~{\rm cm^{-3}}$.
In such high density regions, the neutral fractions can be high owing to the self-shielding effect. However galaxies host stars which produce substantial ionization
photons. 
In general, the neutral fraction in a galaxy is determined by the balance
among the recombination, photo-ionization, and other baryon physical processes.
Therefore, the appropriate evaluation of the neutral fraction in a galaxy may require higher resolution than the Illustris. 
Our result of the 21-cm signal based on the Illustris data has
uncertainty in the contribution from galaxies.
Besides, from the view point to solve the missing baryon problem, it is important to evaluate the 21-signal from the diffuse IGM components in the filamentary structures without the contribution of galaxies. 

To investigate the contribution of the IGM, we evaluate the 21-cm signal from a filament without overdense regions in which galaxies locate in the simulation. 
In Fig.~\ref{fig:angreso}, the blue filled squares are plotted for the case without overdense regions whose 
density contrasts are $\delta>200$~(corresponding to $10^{-3}~\rm
cm^{-3}$ in Fig.~\ref{fig:nxT},
while the green filled circles are for the case
without $\delta >1,000$~(corresponding to $10^{-2}~\rm
cm^{-3}$ in Fig.~\ref{fig:nxT}.
In other words, the blue filled squares show the contribution only from the WHIM in
the filament. 
On the other hand, the red asterisks represent the contribution from both the WHIM and the galaxies and their surroundings, i.e., circumgalactic medium~(CGM).
The comparison with the red asterisks tells us that
most of the 21-cm signals come from the galaxies in the filament.
To resolve the missing baryon problem, the probe of the WHIM in filament structures
is essential. 
However, the 21-cm signal from only the WHIM is too weak for the current SKA design
to detect.
For mapping of the WHIM by 21-cm line observations,
we need not only the removal of the galaxy contribution from the signal map also 10 times better sensitivity than SKA even in large resolutions~($\Delta \theta >20~$arcmin). 

\subsection{Impact of hydrodynamic processes on the 21-cm signal}\label{comparison}

The detectability of the 21-cm signals from the filament structures
has been also evaluated in~\citet{Takeuchi2014}.
In~\citet{Takeuchi2014}, the signals are evaluated from a $N$-body simulation with a simple
thermal evolution depending on the gas number density,
while our signals are based on the
Illustris data in which the coupled equations of gravity and
hydrodynamics are solved with the various feedback effects.  This
difference brings the different distributions of the neutral hydrogen
fraction, and the gap consequently arises in the predicted 21-cm
signals.  In this subsection, we briefly demonstrate how much
the hydrodynamical and feedback effects modify the neutral hydrogen fraction.

\begin{figure}
 \includegraphics[width=\columnwidth]{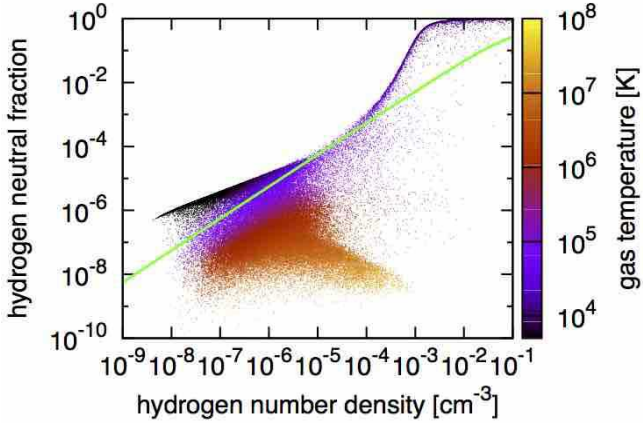}
 \caption{Neutral hydrogen fraction as a function hydrogen number
 density at the present epoch. 
 The coloured dots represent the gas cells in the Illustris simulation, and the color indicating the gas temperature. 
The light green solid line is satisfying the ionization equilibrium with the fixed gas temperature of $3\times 10^{4}~\rm{K}$. }
 \label{fig:nxT}
\end{figure}

The neutral fractions are determined by the net balance between the background photo-ionization rates and the recombination rates as follows;  
\begin{align}
\frac{d}{dt}(x_{\ion{H}{II}})\,\,\,&=\Gamma_{\ion{H}{I}}x_{\ion{H}{I}}-\alpha_{\ion{H}{II}}
n_{\rm{e}}x_{\ion{H}{II}},
\\
\frac{d}{dt}(x_{\ion{He}{II}})\,&=\Gamma_{\ion{He}{I}}x_{\ion{He}{I}}-\Gamma_{\ion{He}{II}}
x_{\ion{He}{II}}\nonumber \\
&\,\,\,\,\,\,-\alpha_{\ion{He}{II}}n_{\rm{e}}x_{\ion{He}{II}}+\alpha_{\ion{He}{III}}n_{\rm{e}}
x_{\ion{He}{III}}, 
\\
\frac{d}{dt}(x_{\ion{He}{III}})&=\Gamma_{\ion{He}{II}}x_{\ion{He}{II}}-\alpha_{\ion{He}{III}}
n_{\rm{e}}x_{\ion{He}{III}}, 
\end{align}
where $\Gamma_{i}$ and $\alpha_{i}$ are the photo-heating rate and the recombination coefficient for the $i$-the species. 
If small amount of metals are negligible, the charge neutrality requests that $n_{\rm{e}}=n_{\ion{H}{II}}+n_{\ion{He}{II}}+2n_{\ion{He}{III}}$. 
Here, we omit the collisional ionization term because the gas temperature is, at most, $T_{\rm K}\sim 10^4$~K without the hydrodynamical and feedback effects, and the collisional ionization effect is negligible at this temperature.
Assuming the ionization equilibrium with $T_{\rm K} = 3 \times 10^4~$K,
we draw a light green line satisfying the relation between the neutral hydrogen fraction $x_{\ion{H}{I}}$ and the hydrogen nucleous number density $n_{\rm H}$ in Fig.~\ref{fig:nxT}. 
The light green line is almost similar to the relation between the neutral hydrogen
fraction and the hydrogen number density in~\citet{Takeuchi2014}. 
Here, to plot the light green line, 
$\Gamma_i$ and $\alpha_i$ are taken from \citet{HM2012} and \citet{FukugitaKawasaki1994}, respectively. 

Fig.~\ref{fig:nxT} obviously shows that the neutral hydrogen fractions
in the Illustris simulation are widely spread rather than the light green
line. This rich distribution produces the difference in the results
between with and without the hydrodynamical and feedback effects.

At high density regions of  $n_{\rm H}>10^{-4}~{\rm cm^{-3}}$, the neutral fractions in the Illustris are higher owing to the self-shielding effect implemented in the Illustris. 
At $10^{-7}~{\rm cm^{-3}}< n_{\rm H} <10^{-5}~{\rm cm^{-3}}$
corresponding the filament regions where we focus on,
the neutral hydrogen fractions in the Illustris
are widely spread around the green light line.
The gas in the filaments is heated by shock and feedback effects, and high collisional ionization rates are expected there,
as shown by previous studies~\citep[e.g.][]{Cen1999, Dave2001,Yoshikawa2003, Bregman2007}.
Therefore, the gas in such regions is distributed below the light green
line in Fig.~\ref{fig:nxT}. On the other hand, the gas in the outer
region of the filament is not heated effectively. 
As a result, the gas at this region is cooled well by the Hubble expansion and the neutral
fraction increases, compared with the light green line. 

\begin{figure}
 \includegraphics[width=\columnwidth]{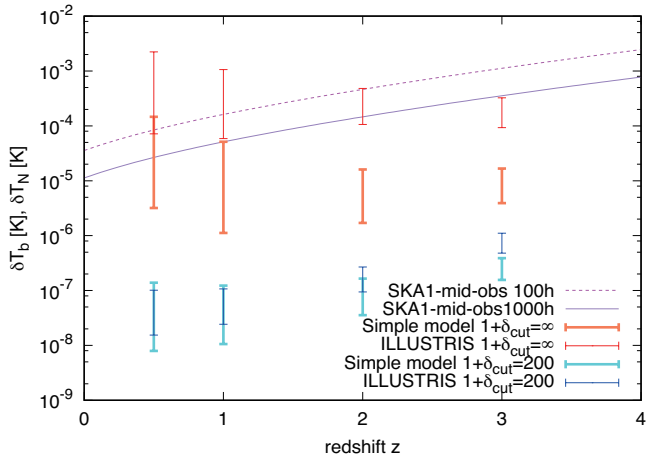}
 \caption{
Redshift evolution of the 21-cm signals in the simple model described in the text. 
The thin orange bar is obtained without the
 density threshold for overdense regions, while the thin light blue
 bar represents the case without the overdense regions with
 $>\delta_{\rm crit} =200$. The thin red and blue bars are for the
 Illustris data with $\delta_{\rm crit} = \infty$ and 200, respectively.  
 The solid line and the
 dashed line are the noise for the total observing time $t_{\rm{obs}}
 = 1,000~\rm{hours}$ and $t_{\rm{obs}} = 100~\rm{hours}$.
 Here, we set the bandwidth $\Delta \nu = 3~\rm{MHz}$ and use $\Delta \theta = 3$ arcmin.
 }\label{fig:compare}
\end{figure}

Fig.~\ref{fig:compare} shows the redshift evolution of the 21-cm signals expected from the simple model described above, i.e.~calculated from the light green line in Fig.~\ref{fig:nxT},
as thick orange and light blue bars.
For comparison, we plot the signals from the Illustris data as thin red
and blue bars.
The simple model does not include contribution from galaxies. 
Therefore, the signal amplitude for the simple model is lower than that for the Illustris simulation in which the galaxy contribution strongly enhances the signal as discussed above.
When we focus on the signals from the WHIM, which are obtained by not
including overdense region with $\delta_{\rm crit} >200$,
the difference between the simple model and the Illustris data becomes small. 
This is because, although Fig.~\ref{fig:nxT} shows that the distribution of the neutral fraction in the filament region is 
widely spread in the Illustris data,
this distribution is smoothed by the spatial resolution of the observation and the signal amplitudes are almost the same in both the cases.
However, at $z=3$, the signal in the Illustris is stronger than that in the simple model. 
At such a high redshift, the structure formation still does not proceed well. 
As a result, the typical temperature of the filamentary gas in the Illustris is lower than $3\times 10^{4}~\rm{K}$ assumed in the simple model, and the resultant high recombination rate increases the neutral hydrogen in the Illustris. 

\section{CONCLUSIONS}\label{CONCLUSIONS} 

The measurement of 21-cm lines due to the hyperfine transition of
neutral hydrogen is a powerful tool to probe the overdense regions in
the IGM. Since they could allow us to survey the regions
which cannot be observed by X-ray and UV/opt telescopes, it is
expected that the observations of 21-cm lines may resolve the so-called
missing baryon problem.
In this paper, focusing the large-scale filamentary structures, we have investigated the detectability of 21-cm signals from the structures by the future radio interferometer telescope, SKA.

First we have generated 21-cm signal maps from the Illustris simulation which is a state-of-the-art cosmological hydrodynamics simulation. 
Using these maps, we have evaluated 21-cm signals from
a filament. 
From the comparison with the noise in the
current design of SKA1-mid, we have investigated the detectability of
these signals at different redshifts. 
Although the detectability depends on the angular resolution of the observation,
we have found that SKA1-mid can detect the signals from the filamentary structure with the total observing time 100--1,000 hours.

We have found that galaxies in the filament provide the significant
contribution to the 21-cm signal, although there still exists theoretical uncertainties in the contribution from galaxies. 
We have also evaluated the detectability of the signals from the WHIM in
the filament structure, removing overdense regions where galaxies exist. 
As a result, it is found that the signals from the WHIM are too small even for the SKA1-mid to detect.
Our result suggests that 10 times better sensitivity than the SKA1-mid is required for probing the WHIM. 
Since, in the evolution of the WHIM, the hydrodynamic and feedback effects are important,
we have investigated the impact of these effects on the signals from the WHIM, comparing the Illustris simulation data with a simple gas model in which the hydrodynamical and feedback effects are neglected. 
The Illustris data show that the values of the neutral hydrogen fraction in the filament
structures are widely spread. 
However, this distribution is smoothed away in the resolution size of
the observation. 
Therefore, the hydrodynamical and feedback effects hardly affect the expected amplitude of the 21-cm signals from the WHIM. 

Although we have concluded that detecting the 21-cm signals from the WHIM is challenging, 
it is promising that SKA1-mid can probe the filament structures at high
redshifts we have not accessed by current galaxy surveys.
The final phase of SKA, SKA2, has much better sensitivity than SKA1-mid.
Therefore, it is expected that SKA2 can survey the detail structures of
the filaments with a better resolution. 
The future 21-cm observation can provide further information on the
structure formation history and the important hints to address the
missing baryon problem.

\section*{Acknowledgements}

We are grateful to N. Sugiyama for valuable comments on our results. 
This work was in par supported by a grant from NAOJ (K.H.), 
Japan Society for the Promotion of Science~(JSPS) KAKENHI Grant 
No.~26-2667 (S.A.),
No.~15K17646 (H.T.) and MEXT's Program for Leading Graduate Schools
``Ph.D. Professionals, Gateway to Success in Frontier Asia'' (H.T.). 

\bibliographystyle{mnras}
\bibliography{MT}


\bsp	
\label{lastpage}
\end{document}